\documentclass[pre,superscriptaddress]{revtex4}
\usepackage{graphicx,amsmath,amsfonts,amssymb}
\begin{document}
\title{Probabilistic ballistic annihilation with continuous 
velocity distributions}
\author{Fran\c{c}ois Coppex}
\affiliation{Department of Physics, University of Gen\`eve, 
CH-1211 Gen\`eve 4, Switzerland}
\author{Michel Droz}
\affiliation{Department of Physics, University of Gen\`eve, 
CH-1211 Gen\`eve 4, Switzerland}
\author{Emmanuel Trizac}
\affiliation{Laboratoire de Physique Th\'eorique (UMR 8627 du CNRS), B\^atiment 210, 
Universit\'e de Paris-Sud, 91405 Orsay, France}
\begin{abstract}
We investigate the problem of ballistically controlled reactions where particles either annihilate upon collision with probability $p$, or undergo an elastic shock with probability $1-p$. Restricting to homogeneous systems, we provide in the scaling regime that emerges in the long time limit, analytical expressions for the exponents describing the time decay of the density and the root-mean-square velocity,  as continuous functions of the probability $p$ and of a parameter related to the dissipation of energy. We work at the level of molecular chaos (non-linear Boltzmann equation), and using a systematic Sonine polynomials expansion of the velocity distribution, we obtain in arbitrary dimension the first non-Gaussian correction and the corresponding expressions for the decay exponents. 
We implement Monte-Carlo simulations in two dimensions, that are in 
excellent agreement with our analytical predictions. 
For $p<1$, numerical simulations lead to conjecture that unlike for pure
annihilation ($p=1$), 
the velocity distribution becomes universal, i.e. does not depend on the initial conditions.
\end{abstract}
\maketitle

%=========================================================
\section{Introduction}\label{section1}
We consider an assembly of particles that move freely in $d$-dimensional space
between collisions, where only two body collisions are taken into account. 
The purpose of this paper is to present a model that unifies both the dynamics of
annihilation \cite{BenNaim,sim2,krapivski,trizac2,trizac} and of hard-sphere 
gases \cite{Resibois} using a continuous parameter $p \in [0,1]$, the probability 
that two particles annihilate when they touch each 
other \cite{Prl}. In the limiting case $p=1$, 
we recover pure annihilation dynamics, and for $p= 0$ the system of hard spheres. 
In our system in the limit $p\to 0$, $p>0$ (denoted $p\to 0^+$), 
a particle will collide elastically many 
times before being annihilated. Thus the particles have a diffusing-like motion 
before annihilating.

Another extensively 
studied
class of problems is the one of diffusion-limited annihilation in which 
diffusing particles annihilate on contact with a given 
rate~\cite{nouveau2,nouveau3,nouveau4}. The simplest case corresponds to the 
reaction $A+A\to \varnothing$. The number of particles decays, in the long time 
regime, as a power law $n(t) \sim t^{-\xi}$. The decay exponent can be exactly 
computed~\cite{michel} and is $\xi = \min (1,d/2)$, where $d$ is the dimension 
of the system. However, the time decay exponents for the density 
found in our case when $p \to 0^+$
are different from the exponents 
found in diffusion-limited systems. The reason for this difference is that the 
underlying microscopic mechanisms responsible for diffusion are different. In our case, particles which have a bigger velocity modulus have a bigger annihilation rate than the slow particles. The velocity dependence of the annihilation rate is not present in the usual diffusion-limited annihilation.

It was recently shown~\cite{trizac} that in the long time limit, the annihilation dynamics for dimensions higher than one is adequately described by the nonlinear Boltzmann equation. This may be understood in a qualitative way by the fact that the density of the gas decays as a function of time, so that the packing fraction (which is the total volume occupied by the particles divided by the total volume of the system) decreases, and the role played by correlations 
(re-collisions) becomes neglectible. The Boltzmann equation thus becomes relevant at late times. With this phenomenology in mind, we conjecture that in the case of probabilistic ballistic annihilation, the Boltzmann equation adequately describes the dynamics
for $p>0$. For $p=0$, the resulting elastic hard sphere system
would be correctly described by Boltzmann's equation in the low density
regime only \cite{lanford,Resibois}.

The paper is organized as follows: in Sec. \ref{section2}, we first introduce the Boltzmann kinetic equation describing the probabilistic annihilation dynamics of a homogeneous system in the scaling regime,
that corresponds to asymptotically large times. We then provide analytical expressions for the exponents $\xi$ and $\gamma$ governing the algebraic time decay of the particle density and the root mean-square-velocity respectively. Next, we give the first non-Gaussian correction $a_2$ to the rescaled velocity distribution by means of a Sonine polynomial expansion. This allows to give explicit expressions for the exponents $\xi$ and $\gamma$ up to the first correction in $a_2$. Sect.~\ref{section3} shows the results of direct Monte-Carlo simulations (DSMC) that are in very good agreement with the analytical results. In the insight of those simulations we clarify the ambiguities following from the analytical computation of $a_2$~\cite{limite} and select the simplest and most accurate relation for $a_2$. It is numerically shown that unlike for pure annihilation, the first Sonine correction for $0 < p < 1$ does not depend on the parameter $\mu$ characterizing the initial distribution $f$ for small velocities: $\lim_{|\mathbf{v}|\to 0} f(\mathbf{v};t=0) \propto |\mathbf{v}|^\mu$. 
We also show analytical and numerical evidence that the conjecture 
put forward in \cite{trizac2} according to which the exponent $\xi = 4 d /(4d+1)$ becomes exact in the limiting case $p\to 0^+$~\cite{trizac2}. 
Finally, Sect. \ref{section4} contains our conclusions.

%=========================================================
\section{Boltzmann kinetic equation}\label{section2}
\subsection{Scaling regime}
We consider a system made of spheres of diameter $\sigma$ moving ballistically in $d$-dimensional space. If two particles touch each other,  they annihilate 
with probability $p$ and thus disappear from the system. 
With probability $1-p$, they undergo an elastic collision. The precollisional velocities 
$\mathbf{v}_i^{**}$ and the postcollisional ones $\mathbf{v}_i$ are related 
in the latter case by
\begin{subequations}
\label{eq1}
\begin{eqnarray}
\mathbf{v}_1^{**} &=& \mathbf{v}_1 - (\mathbf{v}_{12} \cdot \widehat{\boldsymbol{\sigma}})\widehat{\boldsymbol{\sigma}}, \label{eq1a} \\
\mathbf{v}_2^{**} &=& \mathbf{v}_2 + (\mathbf{v}_{12} \cdot \widehat{\boldsymbol{\sigma}})\widehat{\boldsymbol{\sigma}}, 
\label{eq1b}
\end{eqnarray}
\end{subequations}  
where $\mathbf{v}_{12} = \mathbf{v}_1 - \mathbf{v}_2$ is the relative velocity of two particles, and $\widehat{\boldsymbol{\sigma}}$ a unit vector joining the centers of the grains. We consider only two body collisions. 
The initial spatial distribution of particles is supposed to be 
and assumed to remain homogeneous.

Let $f_1(\mathbf{v}_1;t)$ be the instantaneous single particle distribution function in $\mathbb{R}^d$. The Boltzmann equation for our homogeneous system free of forcing reads~\cite{trizac}
\begin{equation}
\frac{\partial}{\partial t} f_1(\mathbf{v}_1;t) = - p \sigma^{d-1} \int d \mathbf{v}_1 d \mathbf{v}_2 d \widehat{\boldsymbol{\sigma}} \, \theta \left( \widehat{\boldsymbol{\sigma}} \cdot \widehat{\mathbf{v}}_{12} \right) \left( \widehat{\boldsymbol{\sigma}} \cdot \mathbf{v}_{12} \right) 
f_1(\mathbf{v}_1;t)f_1(\mathbf{v}_2;t) \\
+ (1-p)\,I_c(f_1,f_1),
\label{boltzcorr} 
\end{equation}
where $\theta$ is the Heaviside distribution, $\widehat{\mathbf{v}}_{12} = \mathbf{v}_{12}/v_{12}$, and $v_{12} = |\mathbf{v}_{12}|$. The integration with respect to $d \widehat{\boldsymbol{\sigma}}$ runs over the solid angle. The first term on the right-hand-side of Eq.~(\ref{boltzcorr}) describes the annihilation dynamics, and the second one the elastic shocks: the collision term $I_c$
reads
\begin{equation}
I_c(f_1,f_1) \,=\,
\sigma^{d-1} \int d \mathbf{v}_1 d \mathbf{v}_2 d \widehat{\boldsymbol{\sigma}} \, \theta \left( \widehat{\boldsymbol{\sigma}} \cdot \widehat{\mathbf{v}}_{12} \right) \left( \widehat{\boldsymbol{\sigma}} \cdot \mathbf{v}_{12} \right) \Big[ 
f_1(\mathbf{v}_1^{**};t) f_1(\mathbf{v}_2^{**};t)-
f_1(\mathbf{v}_1;t) f_1(\mathbf{v}_2;t)
\Big].
\end{equation}

We are searching for an isotropic scaling solution of the homogeneous system, where the time dependence of the distribution function is absorbed into the particles density $n(t)$ and in the typical velocity $\overline{v}(t) = \sqrt{2\langle \mathbf{v}^2 \rangle/d}$, where $\langle v^2\rangle$ is the mean squared velocity. This imposes the scaling form~\cite{trizac,NoijeErnst}
\begin{equation}
f_1(\mathbf{v};t) = \frac{n(t)}{\overline{v}^d(t)}\widetilde{f}(c), \label{scale}
\end{equation}
where the rescaled velocity is given by $c=v/\overline{v}(t)$. 
By construction, $\int \widetilde f=1$.

\subsection{Decay exponents in the scaling regime}\label{decayexp}
By integrating Eq.~(\ref{boltzcorr}) over $\mathbf{v}_1$ with weights $1$ and $v_1^2$, we obtain the number density and energy time evolution
\begin{subequations}
\begin{eqnarray}
\frac{d n}{d t} &=& - p \omega(t) n, \\
\frac{d (n \overline{v}^2)}{d t} &=& - p \alpha \omega(t) n \overline{v}^2,
\end{eqnarray}\label{rate}
\end{subequations}
where the collision frequency $\omega$ is given by
\begin{equation}
\omega(t) = n(t) \overline{v}(t) \sigma^{d-1} \int d\mathbf{c}_1 d \mathbf{c}_2 d \widehat{\boldsymbol{\sigma}} \, \left( \widehat{\boldsymbol{\sigma}} \cdot \mathbf{c}_{12} \right)  \theta \left( \widehat{\boldsymbol{\sigma}} \cdot \mathbf{c}_{12} \right) 
\widetilde{f}(c_1)\widetilde{f}(c_2),
\end{equation}
and the energy dissipation parameter $\alpha$ by
\begin{equation}
\alpha = \frac{\int d\mathbf{c}_1 d \mathbf{c}_2 d \widehat{\boldsymbol{\sigma}} \, \left( \widehat{\boldsymbol{\sigma}} \cdot \mathbf{c}_{12} \right)  \theta \left( \widehat{\boldsymbol{\sigma}} \cdot \mathbf{c}_{12} \right) c_1^2 
\widetilde{f}(c_1)\widetilde{f}(c_2)}{ \left[\int d \mathbf{c} c^2 
\widetilde{f}(c)\right] \left[ \int d\mathbf{c}_1 d \mathbf{c}_2 d \widehat{\boldsymbol{\sigma}} \, \left( \widehat{\boldsymbol{\sigma}} \cdot \mathbf{c}_{12} \right)  \theta \left( \widehat{\boldsymbol{\sigma}} \cdot \mathbf{c}_{12} \right) 
\widetilde{f}(c_1)\widetilde{f}(c_2) \right]}.\label{alpha}
\end{equation}
We made use of the fact that the elastic dynamics does not contribute to the decay of energy or density, thus the integration over the elastic collision term vanishes. The resolution of Eqs.~(\ref{rate}) follows the method 
of Ref.~\cite{trizac} and we obtain
\begin{subequations}
\begin{eqnarray}
\frac{n}{n_0} &=& \left( 1 + p \frac{1+\alpha}{2} \omega_0 t \right)^{-2/(1+\alpha)}, \\
\frac{\overline{v}}{\overline{v}_0} &=& \left( 1+p\frac{1+\alpha}{2}\omega_0 t \right)^{(1-\alpha)/(1+\alpha)},
\end{eqnarray}\label{rateend}
\end{subequations}
where $\omega_0 = \omega(t=0)$ and $\overline{v}_0 = \overline{v}(t=0)$. We conclude from this result that the dynamics are up to the time rescaling $t \to t/p$ (and importantly up to the numerical value of $\alpha$) the same as the ones obtained for pure annihilation~\cite{trizac}. The decay exponents are given by $n(t) \propto t^{-\xi}$ and $\overline{v}(t) \propto t^{-\gamma}$, with
\begin{subequations}
\begin{eqnarray}
\xi &=& \frac{2}{1+\alpha}, \label{decayexponentsa} \\
\gamma &=& \frac{\alpha-1}{\alpha+1}. \label{decayexponentsb}
\end{eqnarray}\label{decayexponents}
\end{subequations}
The scaling exponents consequently satisfy the constraint $\xi + \gamma = 1$.

\subsection{Rescaled kinetic equation}
%-----------------------------------------

Inserting the scaling form (\ref{scale}) in Eq.~(\ref{boltzcorr}) and making use of Eqs.~(\ref{rateend}), we obtain after some algebra
\begin{equation}
\langle c_{12} \rangle \left[ 1 + \frac{1-\alpha}{2}\left( d+c_1 \frac{d}{d c_1} \right) \right] \widetilde{f}(c_1) = \widetilde{f}(c_1)\int_{\mathbb{R}^d} d\mathbf{c}_2 \, |c_{12}| \widetilde{f}(c_2) \\
- \frac{1-p}{p} \frac{1}{\beta_1} \widetilde{I}(\widetilde{f},\widetilde{f}), \label{boltz2}
\end{equation}
where
\begin{equation}
\widetilde{I}(\widetilde{f},\widetilde{f}) = \int_{\mathbb{R}^d} d\mathbf{c}_2 \int d\widehat{\boldsymbol{\sigma}} \, \theta(\widehat{\boldsymbol{\sigma}} \cdot \widehat{\mathbf{c}}_{12}) (\widehat{\boldsymbol{\sigma}} \cdot \mathbf{c}_{12}) \left[ \widetilde{f}(c_1^{**}) \widetilde{f}(c_2^{**}) - \widetilde{f}(c_1)\widetilde{f}(c_2) \right] \label{tildeI}
\end{equation}
and
\begin{equation}
\beta_1 = \int_{\mathbb{R}^d} d\widehat{\boldsymbol{\sigma}} \ \theta(\widehat{\boldsymbol{\sigma}} \cdot \widehat{\mathbf{v}}_{12}) (\widehat{\boldsymbol{\sigma}} \cdot \widehat{\mathbf{v}}_{12}) = 
\frac{\pi^{(d-1)/2}}{\Gamma[(d+1)/2]},
\end{equation}
$\Gamma$ being the gamma function. In equation (\ref{boltz2}), the angular brackets denote average with weight $\widetilde f$: for a given function 
$q(\mathbf{c}_1,\mathbf{c}_2$)
\begin{equation}
\langle q\rangle \,=\, \int d\mathbf{c}_1 d\mathbf{c}_2 \, q(\mathbf{c}_1,\mathbf{c}_2) \, 
\widetilde f(\mathbf{c}_1,\mathbf{c}_2) 
\end{equation}

Making use of the identity~\cite{NoijeErnst}
\begin{equation}
\int_{\mathbb{R}^d} d\mathbf{c} \, c^k \left( d + c \frac{d}{d c} \right) \widetilde{f}(c) = - k \langle c^k \rangle,
\end{equation}
and integrating Eq.~(\ref{boltz2}) over $\mathbf{c}_1$ with weight $c_1^k$, one obtains
\begin{equation}
\alpha = 1 + \frac{2}{k} \left( \frac{\langle c_{12} c_1^k \rangle}{\langle c_{12} \rangle \langle c_1^k \rangle}  -1 \right) + \frac{1-p}{p} \frac{2}{k \beta_1} \frac{\mu_k}{\langle c_{12} \rangle \langle c_1^k \rangle}, \qquad \forall k \geq 0, \label{boltz2b}
\end{equation}
where $\mu_k = - \int_{\mathbb{R}^d} d \mathbf{c}_1 c_1^k \widetilde{I}(\widetilde{f},\widetilde{f})$ and $\alpha = \langle c_{12} c_1^2 \rangle / ( \langle c_{12} \rangle \langle c_1^2 \rangle )$ is the energy dissipation parameter.

\subsection{First non-Gaussian correction}
%--------------------------------------------
The solution of the Boltzmann equation for pure annihilation dynamics 
($p=1$) is non Gaussian in several aspects. The tail of the distribution is overpopulated~\cite{trizac}, and deviations from the Gaussian behavior may also be observed near to the velocity origin~\cite{trizac2,trizac}. It is thus reasonable to think that the velocity distribution function obtained upon solving Eq.~(\ref{boltz2}) will show similar deviations. To study the correction close to the origin, it is convenient to apply a Sonine expansion for the distribution function $\widetilde{f}(c)$ \cite{Landau}
\begin{equation}
\widetilde{f}(c) = \mathcal{M}(c) \bigg[ 1+ \sum_{i\geq 1}a_i S_i(c^2)\bigg], \label{eq7}
\end{equation}
where $\mathcal{M}(c) = \pi^{-d/2} \exp(-c^2)$ is the Maxwellian, and $S_i(c^2)$ the Sonine polynomials. Due to the constraint $\langle c^2\rangle = d/2$, the first correction $a_1$ vanishes~\cite{NoijeErnst}. 
For our purposes, it is sufficient to push the truncation of expression
(\ref{eq7}) to second order, where 
$S_2(x) = x^2/2 - (d+2)x/2+ d(d+2)/8$. 
In order to compute $\alpha$ and $a_2$, one may follow the method used
for inelastic granular gases in Ref. \cite{NoijeErnst}:
we may use the hierarchy (\ref{boltz2b}) for $k=2$ and $k=4$
to obtain a system of two equations for the two unknowns $\alpha$ and $a_2$.
The calculations are however tedious and it appears useful to consider the alternative method that consists in invoking the limit of vanishing velocities of Eq.~(\ref{boltz2})~\cite{trizac}. 
Indeed, since we expect that the tail of the exact solution for the distribution function differs significantly from $\mathcal{M}(c)[ 1+ \sum_{i\geq 1}a_i S_i(c^2)]$, the computation of low order moments of $\widetilde f$ should give a more accurate result. From Eq.~(\ref{eq1})
\begin{equation}
\langle c_{12} \rangle \left[ 1+ d\frac{1-\alpha}{2} \right]\widetilde{f}(0) = \widetilde{f}(0) \langle c_1 \rangle -\frac{1-p}{p} \frac{1}{\beta_1} \lim_{c_1 \to 0} \widetilde{I}(\widetilde{f},\widetilde{f}). \label{boltz3}
\end{equation}
We see that the limit in Eq.~(\ref{boltz3}) involves moments of a lower order than $\mu_4$. Neglecting the corrections $a_i$, $i \geq 3$, the computation of the latter limit gives (see Appendix)
\begin{equation}
\lim_{c_1 \to 0} \widetilde{I}(\widetilde{f},\widetilde{f}) = \frac{\mathcal{S}_d \mathcal{M}(0)}{2 \sqrt{\pi}}\left[ \frac{1-d}{2}a_2 + \frac{d(d+2)}{16}a_2^2 \right], \label{limite}
\end{equation}
where $\mathcal{S}_d = 2 \pi^{d/2}/\Gamma(d/2)$ is the surface of the $d$-dimensional sphere. 
Inserting Eq.~(\ref{limite}) in Eq.~(\ref{boltz3}), 
one obtains a relation between $\alpha$ and $a_2$ that is supplemented with that corresponding to $k=2$ in
(\ref{boltz2b}), in order to finally obtain $\alpha$ and $a_2$. 
To this end, we make use of the various 
relations between moments of the velocity distribution and the fourth 
cumulant $a_2$ derived in \cite{trizac}.
To linear order in $a_2$, the corresponding system reads
\begin{eqnarray}
&&\alpha = 1 + \frac{2}{d} \left( 1- \frac{\sqrt{2}}{2} \right) + a_2 \frac{\sqrt{2}}{2 d} \left[ \frac{1}{8} - \frac{1-p}{p} (d-1) \right]. \label{alpha1}\\
&&\alpha = \frac{\langle c_{12} c_1^2 \rangle}{\langle c_{12} \rangle \langle c_1^2 \rangle} = 1+\frac{1}{2 d} + a_2 \frac{1}{8} \left( 2+ \frac{3}{d} \right) + \mathcal{O}(a_2^2), \label{alpha2}
\end{eqnarray}
where use have been made of the relation $\mu_2=0$ (the elastic shocks
conserve the total kinetic energy of the colliding pairs), which consequently
eliminates $p$ in the second relation.
However, as it was shown in previous works~\cite{limite,Montanero}, there are some ambiguities arising from the linearization procedure, that may affect 
$a_2$ if this quantity is not small enough. We have thus solved the full nonlinear problem, and then in order to have a simpler expression of $a_2$, chosen the linearizing scheme that yields the closest result (the difference does not exceed 10$\%$) to the nonlinear solution. It turns out as well that this scheme is the closest one to the numerical simulations of Sect.~\ref{section3}. This correction is given by:
\begin{equation}
a_2(p) = 8 \frac{3 -2 \sqrt{2}}{4 d + 6 - \sqrt{2} + \frac{1-p}{p} 8 \sqrt{2} (d-1)}. \label{a2p}
\end{equation}
In the limiting case of pure annihilation $p \to 1$, one recovers the result of Ref.~\cite{trizac}.

Inserting this result into the definition Eqs.~(\ref{decayexponents}),
we obtain the decay exponents $\xi$ and $\gamma=1-\xi$. 
In the limit $p\to 0^+$, we note that $a_2$ vanishes, as may have been
anticipated: the velocity distribution then becomes close to its elastic 
Maxwellian counterpart
that holds for $p=0$. In this limit, the decay exponent is $\xi=4d/(4d+1)$,
as conjectured in \cite{trizac2}. We emphasize that the limit
$p\to 0$ is singular: $\xi$ is bounded from above by $4d/(4d+1)$
for any $p>0$, whereas $\xi$ vanishes for $p=0$. It is therefore 
important to exclude $p=0$ from the limit $p\to 0$ in order to
get well behaved limiting expressions.

%=========================================================
\section{Simulation results}\label{section3}

We implement a direct Monte-Carlo simulation (DSMC) scheme in order to solve the Boltzmann equation. The algorithm may briefly be described as follows. We choose at random two different particles $\{i,j\}$. If their velocity is such that $\omega = \mathbf{v}_{ij} \cdot \widehat{\boldsymbol{\sigma}} > 0$, they may collide. Time is subsequently increased by $(N^2 \omega)^{-1}$, where $N$ is the number of remaining particles. With probability $p$ the two particles are then removed from the system, and with probability $1-p$ their velocity is modified according to Eqs.~(\ref{eq1}). For more details on the method see~\cite{trizac,bird,montanero2,frezzotti}. As the number of particles decreases, the statistics at late times suffers from enhanced noise. It is thus necessary to average over many independent realizations.

In dimension one, the dynamics of annihilation creates strong correlations between particles~\cite{sim2}. This precludes a Boltzmann approach that relies on the molecular chaos assumption. We will thus focus on numerical simulations of two-dimensional systems, and we expect the role of correlations to diminish when the dimensionality increases.

\subsection{First Sonine correction}
%-----------------------------------

\begin{figure}
\begin{center}
\includegraphics[width=0.5\columnwidth]{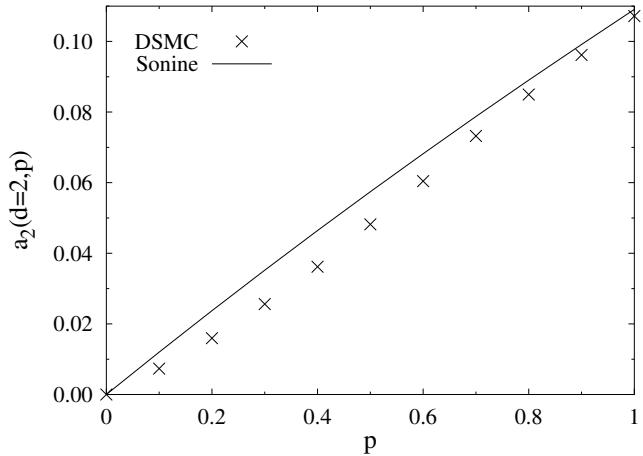}
\end{center}
\caption{First Sonine correction $a_2$ from the analytical estimate~(\ref{a2p}) and from DSMC as a function of the annihilation probability, for $d=2$. The initial number of particles is $5\times 10^6$, and each value is obtained from approximately $10^4$ independent runs. The results are not sensitive to the initial velocity distribution. However, the convergence process is much faster starting from a Gaussian distribution.}
\label{fig1}
\end{figure}

Making use of the relation between $a_2$ and the fourth cumulant of the rescaled velocity distribution~\cite{Montanero}
\begin{equation}
a_2 = \frac{4}{d(d+2)}\langle c^4 \rangle  - 1,
\label{4th}
\end{equation}
we show in Fig.~\ref{fig1} the numerical values of the first Sonine correction $a_2$ for different values of $p$.
The agreement with Eq.~(\ref{a2p}) is good in most cases.

 It turns out that the discrepancy between Eq.~(\ref{a2p}) and DSMC is mainly due to the limit method of computing $a_2$. This method yields a very precise distribution $\widetilde{f}$ in the relevant region of interest in the framework of a Sonine polynomial expansion, namely the small velocity region. On the other hand, it is less accurate in the less interesting high velocity region, hence the discrepancy~\cite{limite}.
\subsection{Decay Exponents}
%-----------------------------------

\begin{figure}
\begin{center}
\includegraphics[width=0.5\columnwidth]{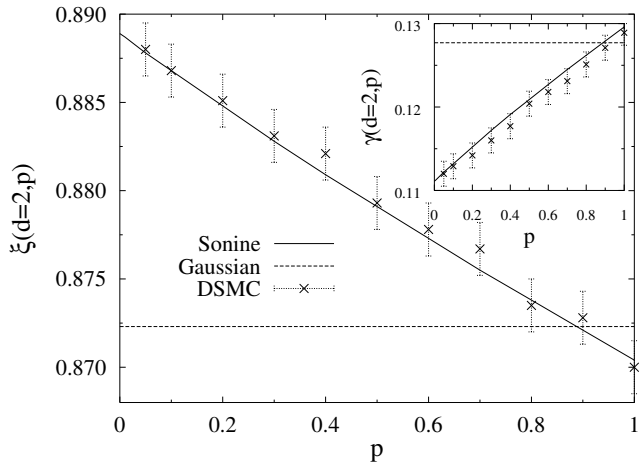}
\end{center}
\caption{The decay exponents $\xi$ and $\gamma$ (inset) in two dimensions, obtained analytically from Eqs.~(\ref{alpha1}) and~(\ref{alpha2}) that are inserted in Eq.~(\ref{decayexponents}), and from DSMC (symbols). The initial number of particles is $5 \times 10^6$, and the number of independent runs approximately $100$. The values of the exponents are not very sensitive to the probability $p$. The horizontal line shows the Maxwellian analytical prediction to zeroth order in $a_2$, i.e. $\xi$ and $\gamma$ from (\ref{alpha1}) and (\ref{decayexponents})
with $a_2=0$.}
\label{fig2}
\end{figure}

\begin{figure}
\begin{center}
\includegraphics[width=0.5\columnwidth]{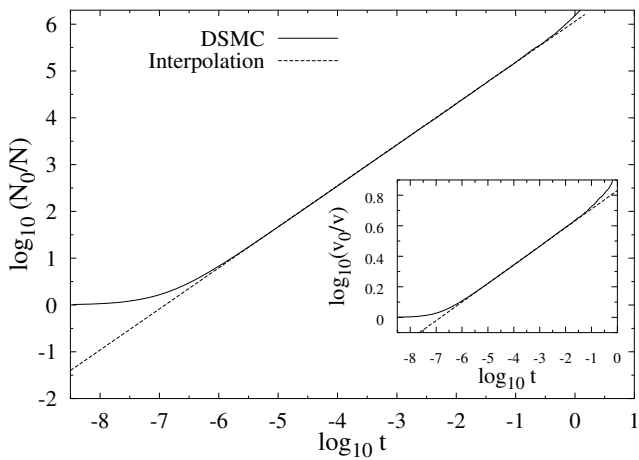}
\end{center}
\caption{The time dependence in two dimensions of $n$ and $\overline{v}$ (inset) on a logarithmic scale for $p=0.6$ and a Gaussian initial velocity distribution, showing a clear power law behavior. The straight line is the linear interpolation giving the decay exponent. $N_0$ ($N$) is the initial (remaining) number of particles. We have denoted $v_0$ the initial root-mean-square velocity $\overline{v}$. The same quantity is denoted $v$ for $t>0$.
The deviation observed for large times is due to the low number of remaining particles.}
\label{fig3}
\end{figure}

\begin{figure}
\begin{center}
\includegraphics[width=0.5\columnwidth]{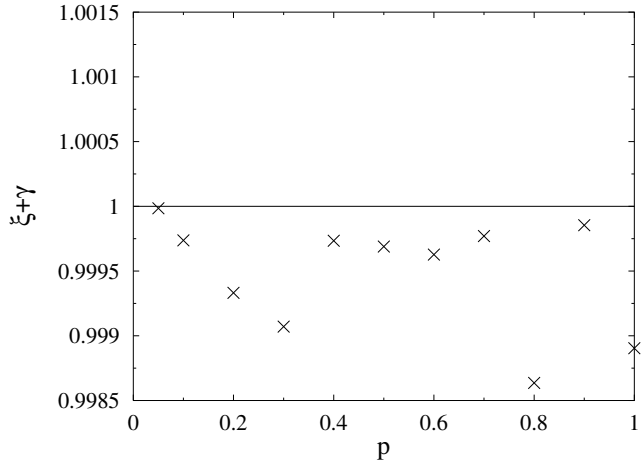}
\end{center}
\caption{Numerical verification of the relation $\xi+\gamma = 1$ in two dimensions for different values of $p$. Note the $y$-scale.}
\label{fig4}
\end{figure}

Plotting the density $n/n_0$ (and the  root-mean-squared velocity $\overline{v}/v_0$) as a function of time $t$ 
on a log-log plot gives the decay exponents (see Fig.~\ref{fig2}).
The numerical results are in agreement with the analytical predictions obtained from the set of Eqs.~(\ref{alpha1}) and~(\ref{alpha2}) that is inserted in Eq.~(\ref{decayexponents}). The predicted power-law behavior is observed over several decades, as shown by Fig.~\ref{fig3} for $p=0.5$.
In Fig.~\ref{fig4}, we show that the scaling relation $\xi + \gamma = 1$ is well obeyed for all values of $p$. Such a relation holds in fact independently of the molecular chaos assumption underlying the Boltzmann equation.

\subsection{Evolution toward the asymptotic distribution}
%--------------------------------------------------------

\begin{figure}
\begin{center}
\includegraphics[width=0.5\columnwidth]{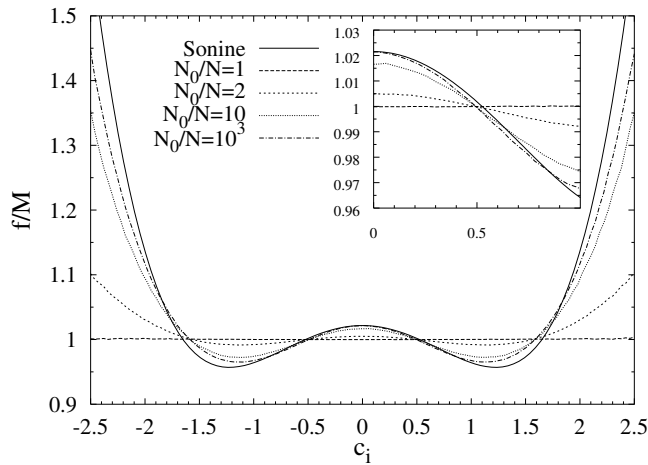}
\end{center}
\caption{Plot of $\widetilde{f}(c_i)/\mathcal{M}(c_1)$ at different times corresponding to different densities, for $p=0.5$. The initial number of particles is $2\times 10^7$ and there are approximately $10^5$ independent runs. The initial distribution is Gaussian and thus corresponds to the flat curve. The continuous curve is the analytical prediction $1+a_2S_2$ with $a_2$ given by Eq.~(\ref{a2p}). The inset shows a magnification of the small velocities region.}
\label{fig6}
\end{figure}

In order to have a more precise understanding and accuracy check of our results, it is useful to study the velocity distribution in the scaling regime. Indeed, the distribution may be adequately described by the Sonine correction $a_2$ at late times only. Before the scaling regime is reached, the velocity distribution $\widetilde{f}(c_1)$ is time-dependent. A very precise check consists in studying the evolution of the non-Gaussianities. To this end, it is useful to consider numerically the quantity $\widetilde{f}(c_i) / \mathcal{M}(c_i) = 1 + a_2 S_2(c_i)$. Fig.~\ref{fig6} shows the evolution of $\widetilde{f}(c_i)/\mathcal{M}(c_i)$ for different times corresponding to different densities, starting from an initial Gaussian distribution.

\begin{figure}
\begin{center}
\includegraphics[width=0.5\columnwidth]{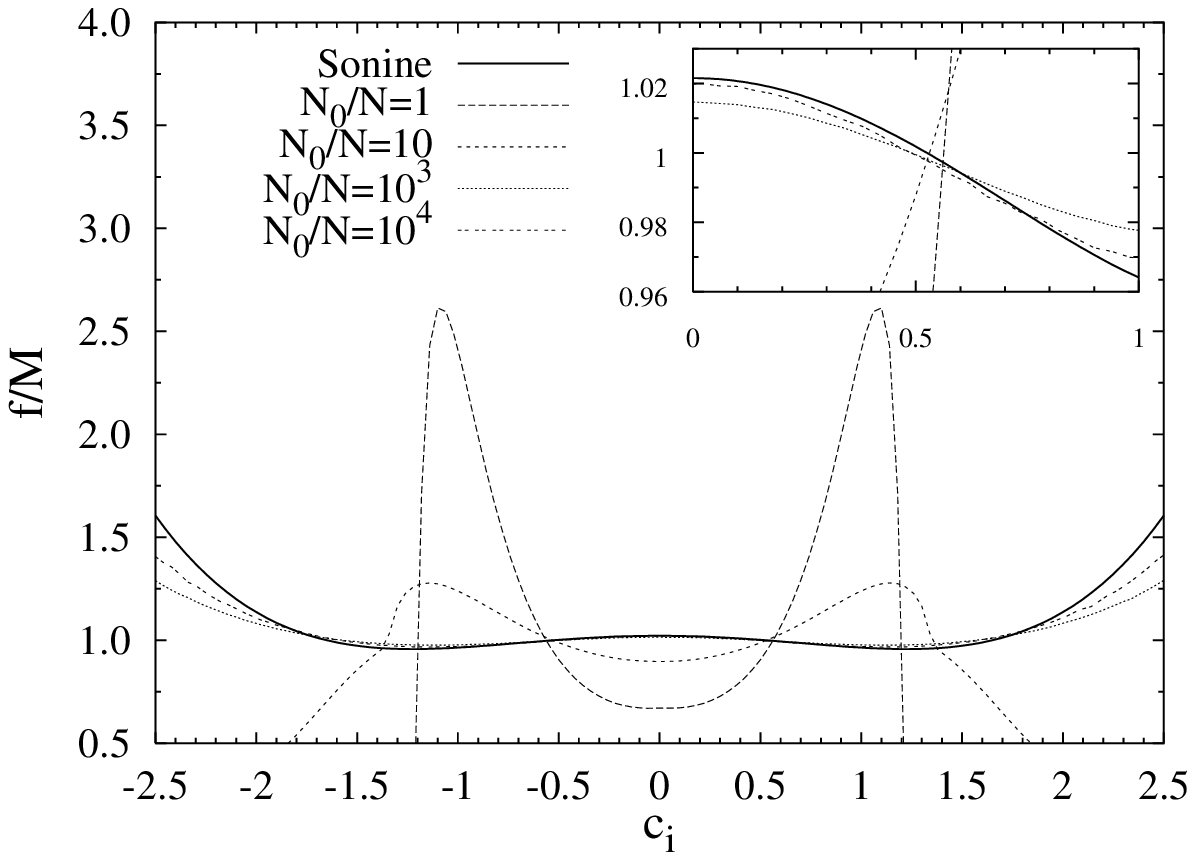}
\end{center}
\caption{Same as Fig.~\ref{fig6} but for an initial distribution such that $\mu=3$.}
\label{fig7}
\end{figure}

\begin{figure}
\begin{center}
\includegraphics[width=0.5\columnwidth]{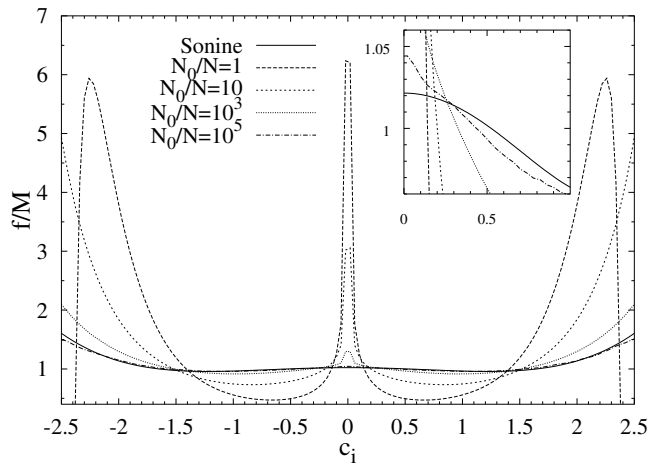}
\end{center}
\caption{Same as Fig.~\ref{fig6} but for an initial distribution such that $\mu=-3/2$ and initially $4\times10^7$ particles.}
\label{fig8}
\end{figure}

It turns out that both methods of computing $a_2$  (directly using its definition in terms of the fourth cumulant (\ref{4th}) or using $\widetilde{f}(c_i)/\mathcal{M}(c_i)$), are fully compatible numerically. However, the latter method requires much more extensive simulations. It is instructive to investigate the evolution toward the asymptotic solution starting from different initial distributions, which are characterized by their behavior near the origin. To this extend we define the exponent $\mu$ by the behavior $\widetilde{f}(c) \simeq |c|^\mu$ for $c \to 0$. Fig.~\ref{fig7} shows the non-Gaussianities of the evolution towards the scaling function for an initial distribution characterized by $\mu = 3$, and Fig.~\ref{fig8} for $\mu = - 3/2$.

\begin{figure}
\begin{center}
\includegraphics[width=0.5\columnwidth]{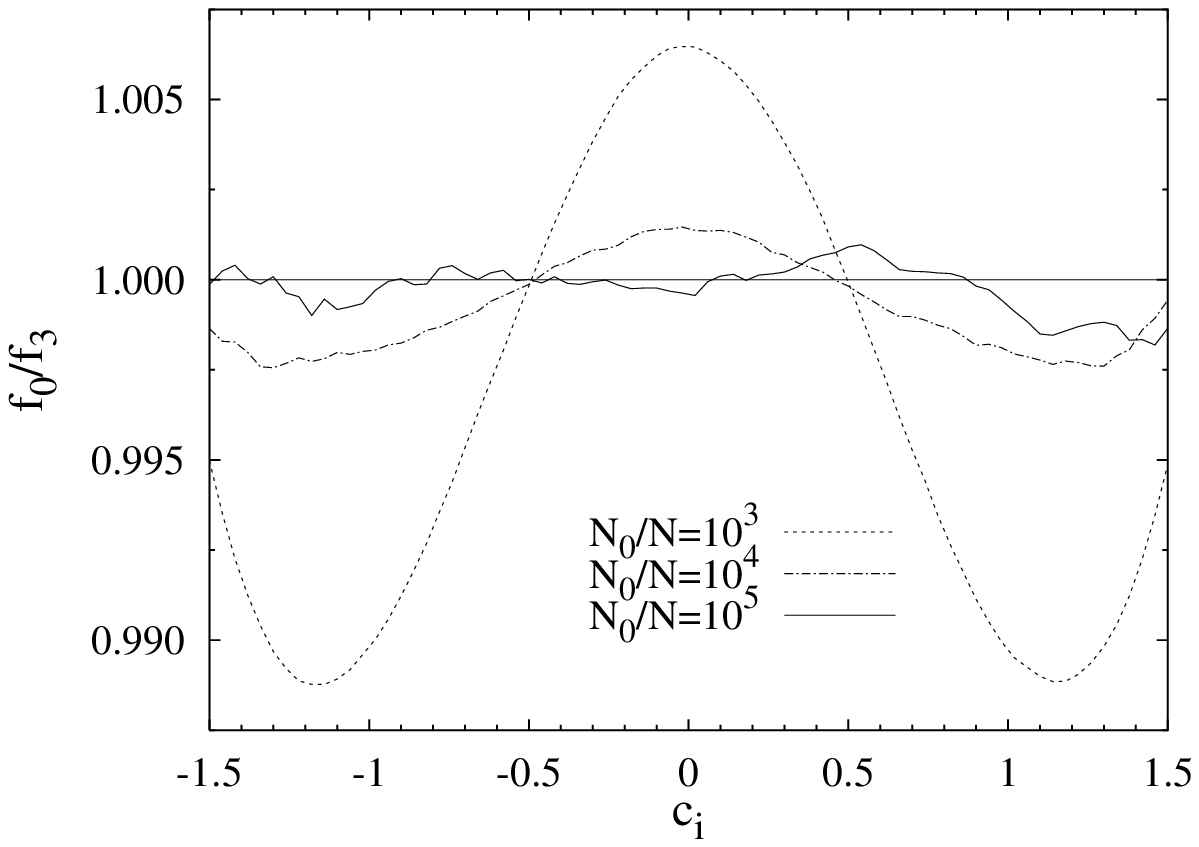}
\end{center}
\caption{Plot of $\widetilde{f}^{(\mu=0)}(c_1)/\widetilde{f}^{(\mu=3)}(c_1)$ for three different times, and $p=0.5$. We see that for late times the ratio of the two distributions tends to unity, which leads to conjecture that the first Sonine corrections $a_2$ are the same in both cases $\mu = 0$ and $\mu = 3$. 
The results reported here correspond to particularly extensive simulations
(note the vertical scale)}
\label{fig9}
\end{figure}

For both initial distributions $\mu = 3$ and $\mu = -3/2$, the solution is attracted toward a scaling function characterized by $\mu = 0$. Hence, there is a qualitative difference between probabilistic annihilation and pure annihilation. Indeed, it was shown in a previous work that for pure annihilation $\mu$ is conserved~\cite{trizac2}, and more importantly that $\mu$ indexes the ``universality classes'' of this process (two distributions with the same $\mu$ 
are characterized by the same long time exponent $\xi$).
 Obviously, adding the effect of elastic collisions in the dynamics of probabilistic annihilation violates the conservation of $\mu$. Next, the question is to know whether the asymptotic distribution depends on $\mu$ or not. We consequently show in Fig.~\ref{fig9} the ratio $\widetilde{f}^{(\mu=0)}(c_1)/\widetilde{f}^{(\mu=3)}(c_1) = (1+a_2^{(\mu=0)})/(1+a_2^{(\mu=3)})$.

The ratio tends to unity, which implies that $a_2^{(\mu=0)} = a_2^{(\mu=3)}$. Moreover, we checked that for the negative value $\mu = - 3/2$, the same conclusion holds. The convergence is however slower due to the divergence of the initial distribution near the velocity origin. We thus conjecture that 
not only the first Sonine coefficient of probabilistic annihilation but
also the full velocity distribution (and hence, all decay exponents)
show an universal property in the sense that they do not depend on the initial velocity distribution if $0<p<1$. This is a nontrivial result since it was shown that this is not true in the case of pure annihilation $p=1$~\cite{trizac2}.

\begin{figure}
\begin{center}
\includegraphics[width=0.5\columnwidth]{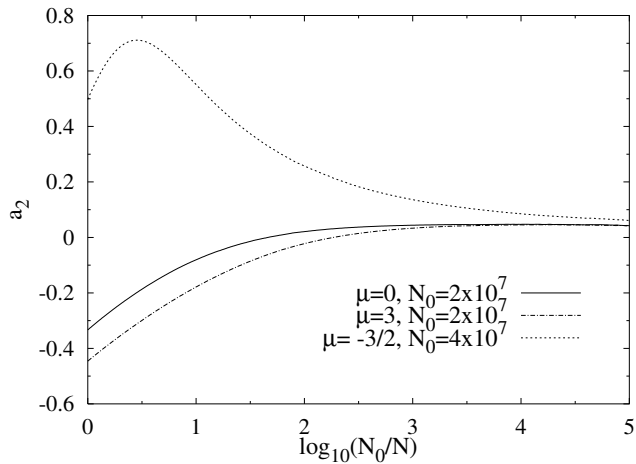}
\end{center}
\caption{Plot of $a_2$ as a function of the densities $N_0/N$ for different values of $\mu$. There are approximately $5\times 10^4$ independent runs.}
\label{fignew}
\end{figure}

Finally, in order to clarify the relevance of the scaling function, 
we studied the fourth cumulant $a_2$ as a function of $N_0/N$, for the same 
parameters as those in Figs.~\ref{fig6}-\ref{fig8}. The result is shown in 
Fig.~\ref{fignew}. The fact that $a_2$ reaches a plateau indicates that the 
system enters a scaling regime at late times. For $\mu=-3/2$ (Fig.~\ref{fig8}), 
due to the initial central peak, the initial distribution is extremely different
from its late time asymptotic counterpart, so that the transient evolution
takes longer and the plateau regime is only approached.
Finally, it may be observed in Fig. \ref{fignew} that for the 3 initial conditions 
the fourth cumulants converge to the same value. This is a further illustration of the
universal behaviour discussed above.

%=========================================================
\section{Conclusions}\label{section4}
A system made of spherical particles moving freely in $d$-dimensional space was studied. When two particles collide, they annihilate with probability $p$ or undergo an elastic collision with probability $1-p$. We gave empirical arguments for the relevance of the Boltzmann description in this system. We obtained analytically the decay exponents of the density of particles and of the root-mean-squared velocity in terms of the energy dissipation parameter $\alpha$. It turns out that upon rescaling time according to $t \to t/p$, $p > 0$, the formal structure of the decay equations is the same as in the case of pure annihilation $p=1$. 

In the scaling regime (that emerges in the long time limit), the first Sonine correction $a_2$ to the Maxwellian distribution was obtained as a function of the continuous parameter $p$. This allows to establish an explicit relation for the decay exponents. It was shown that in the limit $p\to 0^+$, the exponent $\xi$ governing the decay of particles, $n(t) \propto t^{-\xi}$, is given by $\xi = 4d/(4d+1)$, thereby confirming a conjecture put forward in~\cite{trizac2}.

Numerical simulations (DSMC) in two dimensions are in agreement with the analytical correction $a_2(p)$. Moreover, the analytical values for the decay exponents obtained from the first correction $a_2$ are in good agreement as well with numerics. The relation $\xi + \gamma = 1$ was shown to hold for all values of $p$. The study of the dynamics of non-Gaussianities embodied in $a_2 S_2$ reveals a qualitative difference with pure annihilation dynamics: the parameter $\mu$ describing the small velocity behavior of the rescaled distribution is not conserved for probabilistic annihilation when $0<p<1$. Numerical results for different values of $\mu$ leads to conjecture the universality of the rescaled
velocity distribution in this process (this universality being lost
for pure annihilation only, i.e. for $p=1$).

%=========================================================
\begin{acknowledgments} 
We acknowledge useful discussions with J. Piasecki and P. Krapivsky. This work was partially supported by the Swiss National Science Foundation and the French ``Centre National de la Recherche Scientifique''.
\end{acknowledgments}
%=========================================================

\appendix*
\section{Calculation of the limit $c_1 \to 0$ of the collision term
$\widetilde I$}
This quantity may be obtained as a particular case of a previous calculation~\cite{limite}. The result is the following. We define the loss term $\widetilde{I}_l$ and gain term $\widetilde{I}_g$ by
\begin{subequations}
\label{appi}
\begin{eqnarray}
\widetilde{I}_l &=& - \lim_{c_1 \to 0} \int_{\mathbb{R}^d} d\mathbf{c}_2 \int d\widehat{\boldsymbol{\sigma}} \, \theta(\widehat{\boldsymbol{\sigma}} \cdot \widehat{\mathbf{c}}_{12}) (\widehat{\boldsymbol{\sigma}} \cdot \mathbf{c}_{12}) \widetilde{f}(c_1)\widetilde{f}(c_2)\label{appi1}\\
\widetilde{I}_g &=& \lim_{c_1 \to 0} \int_{\mathbb{R}^d} d\mathbf{c}_2 \int d\widehat{\boldsymbol{\sigma}} \, \theta(\widehat{\boldsymbol{\sigma}} \cdot \widehat{\mathbf{c}}_{12}) (\widehat{\boldsymbol{\sigma}} \cdot \mathbf{c}_{12}) \widetilde{f}(c_1^{**}) \widetilde{f}(c_2^{**}),\label{appi2} 
\end{eqnarray}
\end{subequations}
so that $\lim_{c_1 \to 0} \widetilde{I}(\widetilde{f},\widetilde{f}) = \widetilde{I}_l + \widetilde{I}_g$. Within the framework of the Sonine expansion~(\ref{eq7}) and neglecting the coefficients $a_i$, $i \geq 3$, the calculation of the latter integrals gives:
\begin{equation}
\widetilde{I}_l = - \frac{\mathcal{S}_d \mathcal{M}(0)}{2 \sqrt{\pi}} \left[ 1 + a_2 \frac{d(d+2)}{8} \right] \left( 1- \frac{a_2}{8}\right),\label{il}
\end{equation}
\begin{equation}
\widetilde{I}_g = \frac{\mathcal{S}_d \mathcal{M}(0)}{2\sqrt{\pi}} \left[ 1 + a_2 \frac{d^2-2 d + 3}{8} + a_2^2 \frac{3 d(d-2)}{64} \right],\label{ig}
\end{equation}
where $\mathcal{S}_d = 2 \pi^{d/2}/\Gamma(d/2)$ is the surface of the $d$-dimensional sphere, and $\Gamma$ the gamma function. Summing Eqs.~(\ref{il}) and~(\ref{ig}) leads to the result given by Eq.~(\ref{limite}).

%=========================================================

\end{document}